\documentclass[iop]{emulateapj}

\usepackage{natbib,graphicx,amsmath}

\usepackage[usenames]{color}

\shorttitle{Relativistic Gravito-magnetic Instabilities}
\shortauthors{NOH \& HWANG}

\newcommand{\bea}{\begin{eqnarray}}
\newcommand{\eea}{\end{eqnarray}}

\begin{document}

\title{Gravito-magnetic instabilities of Relativistic Magnetohydrodynamics}
\author{Hyerim Noh${}^{1,2}$, Jai-chan Hwang${}^{2,3}$}
\address{${}^{1}$Center for Large Telescope,
         Korea Astronomy and Space Science Institute, Daejon, Korea \\
         ${}^{2}$Kavli Institute for Cosmology Cambridge, Madingley Road, Cambridge CB3 0HA, United Kingdom \\
         ${}^{3}$Department of Astronomy and Atmospheric Sciences,
         Kyungpook National University, Daegu, Korea
         }


\begin{abstract}

We study gravito-magnetic instabilities of a static
homogeneous medium with an aligned magnetic field
in the two contexts of relativistic magnetohydrodynamics (MHD):
first, MHD with post-Newtonian (PN) corrections,
and second, special relativistic (SR) MHD with weak gravity.
The analysis in the PN MHD is made without taking
the temporal gauge condition, thus results are gauge-invariant.
The PN corrections of the internal energy, pressure,
sound velocity and the Alfv\'en velocity lower the critical
(Jeans) wavelength. {All relativistic effects tend to
destabilize the system.}
Although the SR MHD with weak gravity is presented
in the harmonic gauge, in the presence of gravity
the stability analysis is strictly valid to Newtonian order.
In the absence of gravity, the SR MHD is independent of the
gauge condition. We present the plane wave velocities
and the stability criteria in both cases.


\end{abstract}

\maketitle

\section{Introduction}
                                    \label{sec:Introduction}

Recently we presented three formulations of the relativistic
magnetohydrodynamics (MHD):
(i) fully nonlinear and exact perturbation formulation of
MHD in Einstein's gravity,
(ii) special relativistic (SR) MHD with weak (Newtonian) gravity,
and (iii) MHD with first-order post-Newtonian (1PN)
corrections; formulation (ii) is derived in the maximal
slicing (temporal gauge or hypersurface) condition
whereas formulations (i) and (iii) are presented without
imposing the temporal gauge condition
(Noh, Hwang \& Bucher 2019; Hwang \& Noh 2020).
The SR MHD with weak gravity is complementary to the
PN approximation: in the SR MHD with weak gravity the
fluid and field are fully relativistic while the gravity
is Newtonian, whereas in the PN MHD the the fluid, field
and gravity are consistently weakly relativistic.

Previously we studied effects of 1PN corrections on
the MHD waves in a static homogeneous medium with
an aligned magnetic field (Hwang \& Noh 2020).
In this work we include the gravity for the same
homogeneous medium and study the gravito-magnetic
instability of the PN MHD and the SR MHD with weak gravity.
Both formulations include the Newtonian (0PN) limit.

{Gravitational instability is a major factor
causing gravitational collapse to form celestial objects.
Magnetic field ubiquitous in the universe affects the stability.
The gravito-magnetic instability in Newtonian context was analyzed
by Chandrasekhar \& Fermi (1953) and Chandrasekhar (1954, 1961).
Here our aim is to extend this Newtonian study
to a couple of relativistic situations.}

Besides (i) the 1PN {destabilizing} effects on the instability criteria
stated in the abstract and (ii) waves and instability
in the SR MHD with weak gravity, we address
(iii) the issue of inconsistency surrounding the Poisson's
equation in the static homogeneous medium and
(iv) the dependence of instability criteria on magnetic field
to 0PN order.

Sections \ref{sec:PN-MHD} and \ref{sec:SR-MHD}
are summaries of the MHD formulation to 1PN order,
and the SR MHD with weak gravity.
Section \ref{sec:0PN-stability} is the Newtonian (0PN)
study of gravito-magnetic instability of homogeneous
and static medium with a pressure and an aligned magnetic field.
Sections \ref{sec:1PN-stability} and \ref{sec:SR-stability}
are instabilities of the {\it same} medium extended to 1PN order
and the SR MHD with weak gravity, respectively.
Section \ref{sec:Summary} is a summary of our results and
Section \ref{sec:Discussion} is a discussion.
We take the cgs unit.

\section{MHD formulation to 1PN order}
                                    \label{sec:PN-MHD}

We follow 1PN convention of Chandrasekhar
(Chandrasekhar 1965; Chandrasekhar \& Nutku 1969).
The metric convention is
\bea
   & & \widetilde g_{00}
       = - \left\{ 1 - {2 \over c^2} \left[
       U + {1 \over c^2} \left( 2 \Upsilon - U^2 \right)
       \right] \right\},
   \nonumber \\
   & &
       \widetilde g_{0i}
       = - {1 \over c^3} P_i, \quad
       \widetilde g_{ij}
       = \left( 1 + {2 \over c^2} V \right) \delta_{ij},
   \label{metric}
\eea
with $V = U$; $x^0 = ct$
{and a tilde indicates the covariant quantity;
$\Upsilon$ is a pure 1PN order potential
introduced in Chandrasekhar (1965),
see our Equation (\ref{Einstein-1PN-1}).}
The energy momentum tensor is decomposed into fluid quantities as
\bea
   & & \widetilde T_{ab}
       = \widetilde \mu \widetilde u_a \widetilde u_a
       + \widetilde p \left( \widetilde g_{ab}
       + \widetilde u_a \widetilde u_b \right)
       + \widetilde \pi_{ab},
   \\
   & & \widetilde \mu \equiv \mu \equiv \varrho c^2, \quad
       \varrho \equiv \overline \varrho
       \left( 1 + {1 \over c^2} \Pi \right), \quad
       \widetilde p \equiv p, \quad
       \widetilde \pi_{ij} \equiv \Pi_{ij},
   \nonumber \\
\eea
with fluid velocities $v_i$ and $\overline v_i$
\bea
   & & \widetilde u_i \equiv \gamma {v_i \over c}, \quad
       \gamma \equiv {1 \over \sqrt{ 1 - {1 \over 1 + 2 \varphi}
       {v^2 \over c^2}}}, \quad
       v^2 \equiv v^i v_i;
   \nonumber \\
   & &
       {\overline v^i \over c}
       \equiv {\widetilde u^i \over \widetilde u^0}
       = {d x^i \over d x^0}.
\eea
For $\varphi$, see Equation (\ref{metric-FNL}). To 1PN order we have
\bea
   & & v_i = \overline v_i
       + {1 \over c^2} \left(
       3 U \overline v_i
       - P_i \right).
\eea
We will use $\overline v_i$. We are not imposing the temporal gauge condition which can be used as an advantage in handling problems. In our analysis of gravitational instability of PN MHD in Section \ref{sec:1PN-stability}, in fact, we do not need to take the gauge condition. Thus, our results are valid in any temporal gauge condition. 

A complete set of MHD equation valid to 1PN order is derived in Equations (53), (43), (56), (50), (49), (55), (54) and (48), respectively, in Hwang \& Noh (2020) without imposing the temporal gauge condition. The mass, energy and momentum conservation equations are
\bea
   & & {\partial \over \partial t}
       \left[ \overline \varrho \left(
       1 + {1 \over 2} {v^2 \over c^2}
       + 3 {U \over c^2} \right) \right]
   \nonumber \\
   & & \qquad
       + \nabla^i \left[ \overline \varrho \overline v_i
       \left( 1 + {1 \over 2} {v^2 \over c^2}
       + 3 {U \over c^2} \right) \right]
       = 0,
   \label{Mass-conserv-1PN} \\
   & & \overline \varrho \left( \dot \Pi
       + {\bf v} \cdot \nabla \Pi \right)
       + p \nabla \cdot {\bf v}
       + \Pi^{ij} v_{i,j} = 0,
   \label{E-conserv-1PN} \\
   & & \left\{ \overline \varrho
       + {1 \over c^2} \left[ \overline \varrho
       \left( \Pi + v^2 + 6 U \right) + p \right] \right\}
       \left( \dot {\overline v}_i + \overline {\bf v} \cdot \nabla \overline v_i \right)
       + p_{,i}
   \nonumber \\
   & & \qquad
       + \Pi^j_{i,j}
       + {1 \over c^2} \left[
       \dot p v_i
       + 2 p_{,i} U
       + \left( \Pi_{ij} v^j \right)^{\displaystyle{\cdot}}
       - v_i \left( \Pi^j_k v^k \right)_{,j} \right]
   \nonumber \\
   & & \qquad
       - \left\{ \overline \varrho
       + {1 \over c^2} \left[ \overline \varrho
       \left( \Pi + 2 v^2 + 2 U \right) + p \right] \right\}
       U_{,i}
       - {1 \over c^2} 2 \overline \varrho \Upsilon_{,i}
   \nonumber \\
   & & \qquad
       + {1 \over c^2} \overline \varrho
       \left[ \left( 3 \dot U
       + 4 {\bf v} \cdot \nabla U \right) v_i
       - \dot P_i
       - v^j \left( P_{i,j} - P_{j,i} \right) \right]
   \nonumber \\
   & & \qquad
       = {1 \over 4 \pi} \left[
       \left( \nabla \times {\bf B} \right)
       \times {\bf B} \right]_{i}
   \nonumber \\
   & & \qquad
       + {1 \over 4 \pi c^2}
       \Big\{ \left[ \left( {\bf v} \times
       {\bf B} \right)^{\displaystyle{\cdot}} \times {\bf B}
       \right]_i
       + \left( {\bf v} \times {\bf B} \right)_i
       \nabla \cdot \left( {\bf v} \times {\bf B} \right)
   \nonumber \\
   & & \qquad
       + v_i \left( {\bf v} \times {\bf B} \right) \cdot
       \left( \nabla \times {\bf B} \right)
       - \left[ {\bf B} \times \left( {\bf B} \times
       \nabla U \right) \right]_i \Big\}.
   \label{Mom-conserv-1PN}
\eea
Einstein equations are
\bea
   & & \Delta U
       + 4 \pi G \overline \varrho
       = - {1 \over c^2} \bigg[
       2 \Delta \Upsilon
       + 3 \ddot U
       + \dot P^k_{\;\;,k}
   \nonumber \\
   & & \qquad
       + 4 \pi G \overline \varrho \left( \Pi
       + 2 v^2
       + 3 {p \over \overline \varrho }
       + 2 U
       + {B^2 \over 4 \pi \overline \varrho}
       \right)
       \bigg],
   \label{Einstein-1PN-1} \\
   & & \Delta P_i
       - \left( P^k_{\;\;,k}
       + 4 \dot U \right)_{,i}
       = - 16 \pi G \overline \varrho v_i.
   \label{Einstein-1PN-2}
\eea
The Maxwell's equations are
\bea
   & & {\partial \over \partial t}
       \left[ \left( 1 + {U \over c^2} \right)
       {\bf B} \right]
       = \nabla \times
       \left[
       \left( 1 + {U \over c^2} \right)
       \overline {\bf v}
       \times {\bf B} \right].
   \label{Maxwell-1PN-1} \\
   & & \nabla \cdot
       \left[ \left( 1 + {U \over c^2} \right) {\bf B} \right]
       = 0,
   \label{Maxwell-1PN-2}
\eea
Equations (\ref{Mass-conserv-1PN})-(\ref{Maxwell-1PN-2}) are the complete set of MHD equations valid to 1PN order without imposing the temporal gauge condition.

The general slicing condition to the 1PN order is
\bea
   & & P^i_{\;\;,i} + n \dot U = 0,
   \label{Gauge-1PN}
\eea
where $n = 3$ and $4$ correspond to the Chandrasekhar (Standard PN) gauge (Greenberg 1971) and the harmonic gauge (Nazari \& Roshan 2018), respectively (Hwang, Noh \& Puetzfeld 2008; Poisson \& Will 2014).

Equation (\ref{Einstein-1PN-2}) applies only to the 1PN order. Spatial indices of the fluid ($v_i$, $\overline v_i$, $\Pi_{ij}$), field ($B_i$) and the metric ($P_i$) variables are raised and lowered using $\delta_{ij}$ and its inverse metric.

\section{SR MHD with weak gravity formulation}
                                    \label{sec:SR-MHD}

Our metric convention is
\bea
   & & \widetilde g_{00} = - \left( 1 + 2 \alpha \right), \quad
       \widetilde g_{0i} = - \chi_i, \quad
       \widetilde g_{ij} =
       \left( 1 + 2 \varphi \right) \delta_{ij},
   \nonumber \\
   \label{metric-FNL}
\eea
where $\alpha$, $\varphi$ and $\chi_i$ are functions of spacetime with arbitrary amplitudes. Fully nonlinear and exact perturbation formulation based on this metric is presented in the Appendix of Noh, Hwang \& Bucher (2019).

Equations combining SR MHD with weak gravity are derived in Equations (3)-(5) and (10)-(14) in the same work, by {\it assuming}
\bea
   & & \alpha \equiv {\Phi \over c^2} \ll 1, \quad
       \varphi \equiv - {\Psi \over c^2} \ll 1, \quad
       \gamma^2 {t_\ell^2 \over t_g^2} \ll 1,
   \label{WG-conditions}
\eea
where $t_g \sim 1/\sqrt{G \varrho}$
and $t_\ell \sim \ell/c \sim 2 \pi /(kc)$
are gravitational timescale and the light propagation
timescale of a characteristic length scale $\ell$, respectively;
$k$ is the wave number with $\Delta = - k^2$.

The mass, energy, and momentum conservation equations, and the two Maxwell equations, respectively, in conservative forms and Einstein's equations are
\bea
   & & {\partial \over \partial t}
       \left(
       \begin{array}{c}
           D   \\
           E   \\
           m^i \\
           B^i
       \end{array}
       \right)
       + \nabla_j
       \left(
       \begin{array}{c}
           D v^j  \\
           m^j c^2   \\
           m^{ij} \\
           v^j B^i - v^i B^j
       \end{array}
       \right)
   \nonumber \\
   & & \qquad
       =
       \left(
       \begin{array}{c}
           0   \\
           - \overline \varrho \left( 2 \Phi - \Psi \right)_{,i}v^i   \\
           - \overline \varrho \Phi^{,i} \\
           0
       \end{array}
       \right),
   \label{SR-MHD-eqs} \\
   & & B^i_{\;\;,i} = 0,
   \label{Maxwell-eq3} \\
   & & \Delta \Phi
       = 4 \pi G \left( \varrho + {3 p \over c^2}
       + {2 \over c^2} {\cal S} \right)
       = 4 \pi G {E + S \over c^2},
   \label{eq4-SRG} \\
   & & \Delta \Psi
       = 4 \pi G \left( \varrho
       + {1 \over c^2} {\cal S} \right)
       = 4 \pi G {E \over c^2},
   \label{eq2-SRG}
\eea
where
\bea
   & &
       D \equiv \overline \varrho \gamma, \quad
       \varrho \equiv \overline \varrho \left(
       1 + {\Pi \over c^2} \right), \quad
       \gamma
       = {1 \over \sqrt{ 1 - {v^2 \over c^2}}},
   \nonumber \\
   & & E = \varrho c^2 + {\cal S}, \quad
       S = 3 p + {\cal S}, \quad {\cal S} \equiv
       \left( \varrho + {p \over c^2} \right) \gamma^2 {v^2}
   \nonumber \\
   & & \qquad
       + \Pi_{ij} {v^i v^j \over c^2}
       + {1 \over 8 \pi}
       \left[ B^2 + {1 \over c^2}
       \left( {\bf v} \times {\bf B} \right)^2 \right],
   \nonumber \\
   & & m^i \equiv \left( \varrho + {p \over c^2} \right)
       \gamma^2 v^i
   \nonumber \\
   & & \qquad
       + {1 \over c^2} \left[ \Pi^i_j v^j
       + {1 \over 4 \pi} \left( B^2 v^i
       - B^i B^j v_j \right) \right],
   \nonumber \\
   & & m^{ij} \equiv \left( \varrho + {p \over c^2} \right) \gamma^2
       v^i v^j
       + p \delta^{ij}
       + \Pi^{ij}
   \nonumber \\
   & & \qquad
       + {1 \over 4 \pi} \bigg\{
       {1 \over \gamma^2} \left( {1 \over 2} B^2 \delta^{ij}
       - B^i B^j \right)
       + {1 \over c^2} \bigg[ B^2 v^i v^j
   \nonumber \\
   & & \qquad
       + {1 \over 2} \left( B^k v_k \right)^2 \delta^{ij}
       - \left( B^j v^i + B^i v^j \right) B^k v_k
       \bigg] \bigg\}.
\eea
The remaining metric component $\chi_i$ is determined by
\bea
   & & \Delta \chi_i = - {4 \pi G \over c^3}
       \left( 4 \delta_i^j
       - \Delta^{-1} \nabla_i \nabla^j \right) m_j.
   \label{eq3-SRG}
\eea
Indices of $\chi_i$, $m_i$ and $m_{ij}$ are raised and lowered using $\delta_{ij}$ and its inverse.

The SR MHD with weak gravity formulation is valid
in the maximal slicing (the uniform-expansion gauge in cosmology),
setting the trace of extrinsic curvature
(expansion scalar of the normal frame with a minus sign)
equal to zero, which corresponds to the harmonic gauge
in the PN approximation.
If we ignore gravity, the SR MHD is valid in the Minkowski
background, thus independent of the gauge condition.
We have derived the formulation from the fully nonlinear
an exact perturbation formulation of Einstein's gravity
by taking the limit in Equation (\ref{WG-conditions}).
We note that when we consider the conservation equations,
in strict sense of the limit used in
Equation (\ref{WG-conditions}) the gravity part is valid
only to the Newtonian order, thus we have $\Psi = \Phi = -U$
and $\Delta \Phi = 4 \pi G \overline \varrho$
(Noh, Hwang \& Bucher 2019).
We have $v_i = \overline v_i$.

\section{Stability to 0PN order}
                                    \label{sec:0PN-stability}

To the 0PN order, Equations (\ref{Mass-conserv-1PN})-(\ref{Maxwell-1PN-2}) give
\bea
   & & \dot {\overline \varrho}
       + \nabla \cdot \left( \overline \varrho
       {\bf v} \right) = 0,
   \label{Mass-conserv-0PN} \\
   & & \overline \varrho \left( \dot {\bf v}
       + {\bf v} \cdot \nabla {\bf v} \right)
       + \nabla p
       + \nabla_j \Pi^j_i
       - \overline \varrho \nabla U
   \nonumber \\
   & & \qquad
       = {1 \over 4 \pi} \left( \nabla \times
       {\bf B} \right) \times {\bf B},
   \\
   & & \Delta U = - 4 \pi G \overline \varrho,
   \\
   & & \dot {\bf B} = \nabla \times
       \left( {\bf v} \times {\bf B} \right),
   \\
   & & \nabla \cdot {\bf B} = 0.
   \label{Maxwell-2-0PN}
\eea
For our stability analysis we do not need the energy conservation equation in (\ref{E-conserv-1PN}) which is in fact 1PN order (Hwang \& Noh 2013). In this work we {\it ignore} the anisotropic stress. The above equations also follow from Equations (\ref{SR-MHD-eqs})-(\ref{eq2-SRG}) by taking $c$-goes-to-infinite limit with $\Phi = \Psi = - U$.

We {\it consider} a static homogeneous background with a uniform magnetic field (Chandrasekhar \& Fermi 1953; Chandrasekhar 1954, 1961).
To the background order, we have a solution with
\bea
   & & \overline \varrho_0 = {\rm constant}, \quad
       p_0 = {\rm constant}, \quad
       {\bf v}_0 = 0,
   \nonumber \\
   & &
       {\bf B}_0 = B_0 {\bf n} = {\rm constant}, \quad
       |{\bf n}| = 1, \quad
       \nabla U_0 = 0,
   \label{0PN-BG}
\eea
but $\Delta U_0 = - 4 \pi G \overline \varrho_0$.

Notice the inconsistent relations involving $U_0$.
The trouble is caused because we consider Minkowski
(thus static) background despite the presence of self
gravity of an infinite homogeneous background.
The inconsistency is naturally avoided in the relativistic
study of {\it dynamic} background as in the case of cosmology:
{the background order density, $\varrho_0$,
is absorbed into the background (Friedmann) equations, and}
the Poisson's equation is valid only to perturbed order,
thus $U = \delta U$ without $U_0$,
see Equations (85)-(88), (108) and (119) in
Hwang, Noh \& Puetzfeld (2008).
In the static background, as in the present case,
however, the inconsistency (often known as a swindle) remains.
Jeans has made an explicit choice of ignoring the inconsistency,
see Section 46 of Jeans (1902).
Although the $U_0$ term does not appear in perturbation
analysis to 0PN order, it appears in the perturbation
equations to the 1PN order,
see Equations (\ref{Mass-conserv-PN})-(\ref{Gauge-PN}).
See a paragraph below Equation (\ref{Gauge-PN})
for further discussion.

Considering the linear order perturbation, we expand
\bea
   & & \overline \varrho = \overline \varrho_0
       \left( 1 + \delta \right), \quad
       p = p_0 + \delta p, \quad
       {\bf B} = {\bf B}_0 + \delta {\bf B},
   \nonumber \\
   & &
       {\bf v} = \delta {\bf v}, \quad
       U = U_0 + \delta U,
\eea
and introduce
\bea
   & & \delta p \equiv c_s^2 \overline \varrho_0 \delta, \quad
       c_A^2 \equiv {B_0^2 \over 4 \pi \overline \varrho_0}, \quad
       \delta {\bf B} \equiv B_0 {\bf b},
\eea
with the adiabatic sound speed $c_s = {\rm constant}$;
{thus, we are considering a barotropic equation of state
with $p = p(\overline \varrho)$.}
Equations (\ref{Mass-conserv-0PN})-(\ref{Maxwell-2-0PN}) give
\bea
   & & \dot \delta = - \nabla \cdot {\bf v},
   \label{Mass-conserv-0PN-linear} \\
   & & \dot {\bf v}
       + c_s^2 \nabla \delta
       - \nabla \delta U
       = c_A^2 \left[ \nabla \left( {\bf n} \cdot {\bf b} \right)
       - {\bf n} \cdot \nabla {\bf b} \right],
   \\
   & & \Delta \delta U = - 4 \pi G \overline \varrho_0 \delta,
   \\
   & & \dot {\bf b}
       = {\bf n} \cdot \nabla {\bf v}
       - {\bf n} \nabla \cdot {\bf v},
   \label{Maxwell-1-0PN-linear} \\
   & & \nabla \cdot {\bf b} = 0.
   \label{Maxwell-2-0PN-linear}
\eea
Expanding the perturbations in plane waves proportional to $e^{ i({\bf k} \cdot {\bf x} - \omega t)}$, we have
\bea
   & & \left[ \omega^2
       - c_A^2 \left( {\bf n} \cdot {\bf k} \right)^2
       \right] {\bf v}
       + \bigg[ \left( {4 \pi G \overline \varrho_0 \over k^2}
       - c_s^2 - c_A^2 \right) {\bf k}
   \nonumber \\
   & & \qquad
       + c_A^2 {\bf n} {\bf n} \cdot {\bf k} \bigg]
       {\bf k} \cdot {\bf v}
       + c_A^2 {\bf k} {\bf k} \cdot {\bf n}
       {\bf n} \cdot {\bf v}
       = 0.
   \label{v-eq-0PN}
\eea

By introducing coordinates as (Shu 1992)
\bea
   & & {\bf k} \equiv k \widehat {\bf x}, \quad
       {\bf n} \equiv \cos{\psi} \widehat {\bf x}
       + \sin{\psi} \widehat {\bf y}, \quad
       {\bf v} \equiv v_x \widehat {\bf x}
       + v_y \widehat {\bf y} + v_z \widehat {\bf z},
   \nonumber \\
   \label{coordinate}
\eea
Equations (\ref{Mass-conserv-0PN-linear}), (\ref{Maxwell-1-0PN-linear}) and (\ref{Maxwell-2-0PN-linear}) give $v_x = (\omega/k) \delta$, $b_x = 0$,
\bea
   & & b_y = {k \over \omega} \left( \sin{\psi} v_x
       - \cos{\psi} v_y \right), \quad
       b_z = - {k \over \omega} \cos{\psi} v_z,
\eea
and Equation (\ref{v-eq-0PN}) gives
\bea
   & & \left[ \omega^2 + k^2 \left(
       {4 \pi G \overline \varrho_0 \over k^2}
       - c_s^2 - c_A^2 \sin^2{\psi} \right) \right] v_x
   \nonumber \\
   & & \qquad
       + k^2 c_A^2 \sin{\psi} \cos{\psi} v_y = 0,
   \label{x-comp} \\
   & & k^2 c_A^2 \cos{\psi} \sin{\psi} v_x
       + \left( \omega^2
       - k^2 c_A^2 \cos^2{\psi} \right) v_y = 0,
   \label{y-comp} \\
   & & \left( \omega^2 - k^2 c_A^2 \cos^2{\psi} \right) v_z = 0.
   \label{z-comp}
\eea

For $v_z \neq 0$ (${\bf v}$ perpendicular to ${\bf B}_0$-${\bf k}$ plane), Equation (\ref{z-comp}) gives
\bea
   & & {\omega^2 \over k^2}
       = c_A^2 \cos^2{\psi}.
   \label{z-comp-PN}
\eea

\begin{figure}[h]
\hspace{-.3cm} \vspace{-.5cm}
\includegraphics[width=65mm,angle=90]{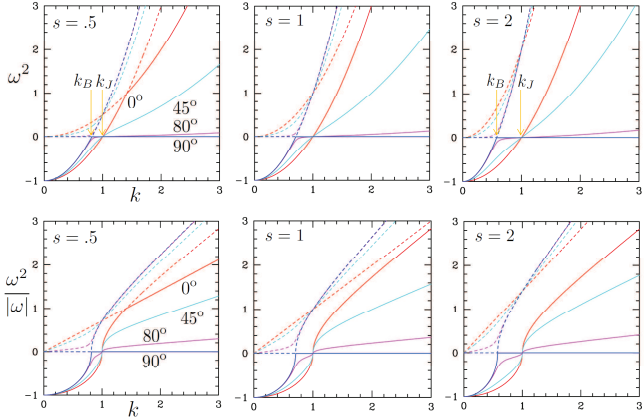}
\caption{Behaviors of two solutions in Equation (\ref{omega-sol})
with the upper (plus) and lower (minus) signs
in dotted and solid lines, respectively.
The $x$-axis is $k$ in the unit of $k_J$ and the $y$-axis
is $\omega^2$ in the unit of $4 \pi G \overline \varrho_0$.
The red, cyan, magenta and blue indicate $\psi = 0^{\rm o}$,
$45^{\rm o}$, $80^{\rm o}$ and $90^{\rm o}$, respectively.
From left to right are for $s \equiv c_A^2/c_s^2 = 0.5$,
$1$ and $2$, respectively.
The lower graphs are the same behaviors using
$\omega^2/|\omega|$ for $y$-axis; these correspond to
Figure 136 in Chandrasekhar (1961).}
 \label{fig-BG}
\end{figure}

For non-vanishing $v_x$ and $v_y$ (${\bf v}$ in ${\bf B}_0$-${\bf k}$ plane), from Equations (\ref{x-comp}) and (\ref{y-comp}) we have a dispersion relation
\bea
   & & \omega^4 + \left[ 4 \pi G \overline \varrho_0
       - k^2 \left( c_s^2 + c_A^2 \right) \right] \omega^2
   \nonumber \\
   & & \qquad
       + c_A^2 \left( k^2 c_s^2 - 4 \pi G \overline \varrho_0
       \right) \left( {\bf k} \cdot {\bf n} \right)^2
       = 0,
   \label{0PN-dispersion}
\eea
with two solutions
\bea
   & & {\omega^2 \over k^2}
       = {1 \over 2} \bigg\{ c_s^2 + c_A^2
       - {4 \pi G \overline \varrho_0 \over k^2}
       \pm \bigg[ \left( c_s^2 + c_A^2 -
       {4 \pi G \overline \varrho_0 \over k^2} \right)^2
   \nonumber \\
   & & \qquad
       - 4 c_A^2 \left( c_s^2
       - {4 \pi G \overline \varrho_0 \over k^2} \right)
       {( {\bf k} \cdot {\bf n} )^2 \over k^2}
       \bigg]^{1/2} \bigg\}.
   \label{omega-sol}
\eea
Equation (\ref{0PN-dispersion}) was presented in
Equation (169) of Chandrasekhar \& Fermi (1953).
{Behavior of the solutions are presented in
Figure \ref{fig-BG}.}

\vspace{-.3cm}
\vspace{-.3cm}

For ${\bf k} \parallel {\bf B}_0$ ($\psi = 0^{\rm o}$), we have
\bea
   & & {\omega^2 \over k^2}
       = c_s^2 - {4 \pi G \overline \varrho_0 \over k^2}, \quad
       c_A^2,
\eea
with the faster (slower) mode the fast (slow) MHD waves in the absence of gravity (Shu 1992). In the presence of gravity one mode is unstable for $k < k_J$ with the Jeans wavenumber given as
\bea
   & & k^2_{J} = {4 \pi G \overline \varrho_0 \over c_s^2},
   \label{Jeans}
\eea
which does not depend on the magnetic field as pointed out in Chandrasekhar \& Fermi (1953); for example, Chandrasekhar (1954) mentions ``Jeans's criterion for the gravitational instability of an infinite homogeneous medium is unaffected by the presence of a magnetic field."; however, this is not a generally valid conclusion as we show below.

For ${\bf k} \perp {\bf B}_0$ ($\psi = 90^{\rm o}$), we have
\bea
   & & {\omega^2 \over k^2}
       = \left( c_s^2 + c_A^2 \right)
       - {4 \pi G \overline \varrho_0 \over k^2}, \quad
       0,
\eea
with the fast mode the magnetosonic wave in the absence of gravity (Shu 1992) and the slow mode vanishing. In the presence of gravity one mode is unstable for $k < k_B$ with the critical wavenumber modified as
\bea
   & & k^2_{B} = {4 \pi G \overline \varrho_0
       \over c_s^2 + c_A^2},
   \label{Jeans-B}
\eea
thus depends on the magnetic field with $k_B < k_J$
(Pacholczyk \& Stod\'o{\l}kiewicz 1960; {Strittmatter 1966}).

For $\psi$ other than $90^{\rm o}$, the stability criterion
remains the same as the Jeans criterion in Equation (\ref{Jeans}).
However, as Figure \ref{fig-BG} shows, the instability
for $k_B < k <k_J$ is suppressed depending on $\psi$,
and as $\psi$ approaches $90^{\rm o}$ the stability
criterion effectively becomes Equation (\ref{Jeans-B}).

\section{Stability to 1PN order}
                                    \label{sec:1PN-stability}

We similarly consider a static homogeneous background without anisotropic stress. To the background order, we have a solution with
\bea
   & & \overline \varrho_0 = {\rm constant}, \quad
       \Pi_0 = {\rm constant}, \quad
       p_0 = {\rm constant},
   \nonumber \\
   & &
       {\bf v}_0 = 0 = \overline {\bf v}_0, \quad
       {\bf B}_0 = B_0 {\bf n} = {\rm constant},
   \nonumber \\
   & &
       \nabla U_0 = 0 = \dot U_0, \quad
       \nabla \Upsilon_0 = 0, \quad
       {\bf P}_{0} = 0,
\eea
but we have
\bea
   & & \Delta U_0 + 4 \pi G \overline \varrho_0
       = - {1 \over c^2} \big[ 2 \Delta \Upsilon_0
   \nonumber \\
   & & \qquad
       + 4 \pi G \overline \varrho_0
       \left( \Pi_0 + 2 U_0 \right)
       + 12 \pi G p_0
       + G B_0^2 \big],
   \label{Poisson-BG}
\eea
with $\Delta U_0 \neq 0 \neq \Delta \Upsilon_0$. Thus, the inconsistent relations involving $U_0$ and $\Upsilon_0$ continue to 1PN order, see a paragraph below Equation (\ref{Gauge-PN}). We consider $p_0$ and $\Pi_0$ as constants in space and time.

To the linear order perturbation, we expand
\bea
   & & \overline \varrho = \overline \varrho_0
       \left( 1 + \delta \right), \quad
       \Pi = \Pi_0 + \delta \Pi, \quad
       p = p_0 + \delta p,
   \nonumber \\
   & &
       {\bf B} = {\bf B}_0 + \delta {\bf B}, \quad
       {\bf v} = \delta {\bf v}, \quad
       \overline {\bf v} = \delta \overline {\bf v},
   \nonumber \\
   & &
       U = U_0 + \delta U, \quad
       \Upsilon = \Upsilon_0 + \delta \Upsilon, \quad
       {\bf P} = \delta {\bf P}.
\eea
Equations (\ref{Mass-conserv-1PN})-(\ref{Gauge-1PN}) give
\bea
   & & \dot \delta = - \nabla \cdot \overline {\bf v}
       - {3 \over c^2} \delta \dot U,
   \label{Mass-conserv-PN} \\
   & & \delta \dot \Pi
       = - {p_0 \over \overline \varrho_0}
       \nabla \cdot {\bf v},
   \label{E-conserv-PN} \\
   & & \left[ 1 + {1 \over c^2} \left( \Pi_0
       + {p_0 \over \overline \varrho_0} + 6 U_0 \right) \right]
       \dot {\overline {\bf v}}
       + \left( 1 + {2 \over c^2} U_0 \right)
       c_s^2 \nabla \delta
   \nonumber \\
   & & \qquad
       - \left[ 1 + {1 \over c^2} \left( \Pi_0
       + {p_0 \over \overline \varrho_0} + 2 U_0 \right) \right]
       \nabla \delta U
   \nonumber \\
   & & \qquad
       - {1 \over c^2} \left(
       2 \nabla \delta \Upsilon
       + \dot {\bf P} \right)
   \nonumber \\
   & & \qquad
       = c_A^2 \left[
       - \nabla \left( {\bf n} \cdot {\bf b} \right)
       + {\bf n} \cdot \nabla {\bf b}
       \right]
   \nonumber \\
   & & \qquad \qquad
       + {c_A^2 \over c^2}
       \left[
       - \left( \dot {\bf v} - \nabla \delta U \right)
       + {\bf n} {\bf n} \cdot \left( \dot {\bf v}
       - \nabla \delta U \right)
       \right],
   \label{Mom-conserv-PN} \\
   & & \Delta \delta U + 4 \pi G \overline \varrho_0 \delta
       = - {1 \over c^2} \Big\{
       2 \Delta \delta \Upsilon
   \nonumber \\
   & & \qquad
       + 4 \pi G \overline \varrho_0
       \left[ \delta \Pi + 2 \delta U
       + \left( \Pi_0 + 2 U_0 \right) \delta \right]
       + 12 \pi G \delta p
   \nonumber \\
   & & \qquad
       + 2 G B_0^2 {\bf n} \cdot {\bf b}
       + 3 \delta \ddot U
       + \nabla \cdot \dot {\bf P} \Big\},
   \label{E-constr-PN} \\
   & & \Delta {\bf P}
       - \nabla \left( \nabla \cdot {\bf P}
       + 4 \delta \dot U \right)
       = - 16 \pi G \overline \varrho_0 {\bf v},
   \label{Mom-constr-PN} \\
   & & \dot {\bf b}
       = {\bf n} \cdot \nabla \overline {\bf v}
       - {\bf n} \nabla \cdot \overline {\bf v}
       - {1 \over c^2} {\bf n} \delta \dot U,
   \label{Maxwell-1-PN} \\
   & & \nabla \cdot {\bf b}
       = - {1 \over c^2} {\bf n}
       \cdot \nabla \delta U,
   \label{Maxwell-2-PN} \\
   & & \nabla \cdot {\bf P} = - n \delta \dot U.
   \label{Gauge-PN}
\eea

In Equations (\ref{Mass-conserv-PN})-(\ref{Gauge-PN})
the $U_0$ term appears in the 1PN order.
To the 0PN order, as the $U_0$ does {\it not} appear
in the stability analysis we can ignore the Poisson's equation
in the background order, assuming the Poisson's equation
valid only to the perturbed order (Jeans 1902).
However, situation becomes more ambiguous to the 1PN order
as the $U_0$, if we keep it, appears directly in
the perturbation equations. As mentioned in a paragraph
below Equation (\ref{0PN-BG}), in the relativistic study
as in cosmology, the background is governed by dynamic
equations like the Friedmann equations, and we have
$U =\delta U$, thus $U_0 \equiv 0$ (Hwang, Noh \& Puetzfeld 2008).
However, in a static (Minkowski) background the inconsistency
cannot be resolved. In the following we will keep
track of $U_0$ term in our analysis so that we can
either ignore the $U_0$ term (Jeans' choice) or use
Equation (\ref{Poisson-BG}) to 0PN order.

We expand the perturbations in plane waves proportional
to $e^{i({\bf k} \cdot {\bf x} - \omega t)}$.
Equations (\ref{Mass-conserv-PN})-(\ref{Maxwell-2-PN}) give
\begin{widetext}
\bea
   & & \left\{ \left[ 1 + {1 \over c^2} \left( \Pi_0
       + {p_0 \over \overline \varrho_0} + 6 U_0
       - 4 {4 \pi G \overline \varrho_0 \over k^2}
       + c_A^2 \right) \right]
       \omega^2
       - c_A^2 \left( {\bf n} \cdot {\bf k} \right)^2
       \right\}
       {\overline {\bf v}}
       + \bigg\{
       - c_s^2 \left[ 1 + {2 \over c^2}
       \left( U_0
       - 3 {4 \pi G \overline \varrho_0 \over k^2} \right) \right]
   \nonumber \\
   & & \qquad
       - c_A^2 \left( 1 - {4 \over c^2}
       {4 \pi G \overline \varrho_0 \over k^2} \right)
       + {4 \pi G \overline \varrho_0 \over k^2}
       \left[ 1 + {1 \over c^2} \left( 2 \Pi_0
       + 2 {p_0 \over \overline \varrho_0}
       + 4 U_0
       - {4 \pi G \overline \varrho_0 \over k^2}
       + {\omega^2 \over k^2} \right) \right] \bigg\}
       {\bf k}
       {\bf k} \cdot \overline {\bf v}
   \nonumber \\
   & & \qquad
       + \left( 1 - {2 \over c^2}
       {4 \pi G \overline \varrho_0 \over k^2} \right)
       c_A^2 {\bf n} {\bf n} \cdot {\bf k}
       {\bf k} \cdot \overline {\bf v}
       + c_A^2 \left[ \left( 1 - {2 \over c^2}
       {4 \pi G \overline \varrho_0 \over k^2} \right)
       {\bf k} {\bf k} \cdot {\bf n}
       - {1 \over c^2} \omega^2 {\bf n} \right]
       {\bf n} \cdot \overline {\bf v}
       = 0.
   \label{v-2}
\eea
{To derive this we start from
Equation (\ref{Mom-conserv-PN}) for $\overline {\bf v}$ and
replace all the other variables in terms of $\overline {\bf v}$
using the perturbation expansion to 1PN order.
Notice that we have {\it not} imposed
the gauge condition in Equation (\ref{Gauge-PN}):
i.e., no choice of $n$ is needed for our stability analysis.
Thus, our result in this section is valid independently
of the temporal gauge condition.}
By taking a coordinate in Equation (\ref{coordinate}),
Equation (\ref{v-2}) gives
\bea
   & & \bigg\{ \left[ 1 + {1 \over c^2}
       \left( \Pi_0 + {p_0 \over \overline \varrho_0}
       + 6 U_0
       - 3 {4 \pi G \overline \varrho_0 \over k^2}
       + c_A^2 \sin^2{\psi} \right) \right] \omega^2
       - \left[ 1 + {2 \over c^2}
       \left( U_0
       - 3 {4 \pi G \overline \varrho_0 \over k^2} \right)
       \right] k^2 c_s^2
   \nonumber \\
   & & \qquad
       - \left( 1 - {4 \over c^2}
       {4 \pi G \overline \varrho_0 \over k^2} \right)
       k^2 c_A^2 \sin^2{\psi}
       + \left[ 1 + {1 \over c^2}
       \left( 2 \Pi_0 + 2 {p_0 \over \overline \varrho_0}
       + 4 U_0
       - {4 \pi G \overline \varrho_0 \over k^2}
       \right) \right] {4 \pi G \overline \varrho_0 \over k^2} k^2 \bigg\} v_x
   \nonumber \\
   & & \qquad
       + \left[ \left( 1 - {2 \over c^2}
       {4 \pi G \overline \varrho_0 \over k^2} \right) k^2
       - {1 \over c^2} \omega^2 \right] c_A^2 \sin{\psi}
       \cos{\psi} v_y = 0,
   \label{x-comp-PN} \\
   & & \left[ \left( 1 - {2 \over c^2}
       {4 \pi G \overline \varrho_0 \over k^2} \right) k^2
       - {1 \over c^2} \omega^2 \right] c_A^2 \sin{\psi}
       \cos{\psi} v_x
   \nonumber \\
   & & \qquad
       + \left\{ \left[ 1 + {1 \over c^2}
       \left( \Pi_0 + {p_0 \over \overline \varrho_0}
       + 6 U_0
       - 4 {4 \pi G \overline \varrho_0 \over k^2}
       + c_A^2 \cos^2{\psi} \right) \right] \omega^2
       - c_A^2 k^2 \cos^2{\psi} \right\} v_y
       = 0,
   \label{y-comp-PN} \\
   & & \left\{
       \left[ 1 + {1 \over c^2}
       \left( \Pi_0 + {p_0 \over \overline \varrho_0}
       + 6 U_0
       - 4 {4 \pi G \overline \varrho_0 \over k^2}
       + c_A^2 \right) \right] \omega^2
       - c_A^2 k^2 \cos^2{\psi} \right\} v_z = 0,
   \label{z-comp-PN}
\eea
with ${\bf k} \cdot {\bf n} = k \cos{\psi}$.

For $v_z \neq 0$ (${\bf v}$ perpendicular to ${\bf B}_0$-${\bf k}$ plane), Equation (\ref{z-comp-PN}) gives
\bea
   & & {\omega^2 \over k^2}
       = c_A^2 \cos^2{\psi}
       \left[ 1 - {1 \over c^2}
       \left( \Pi_0 + {p_0 \over \overline \varrho_0}
       + c_A^2
       - 4 {4 \pi G \overline \varrho_0 \over k^2}
       + 6 U_0
       \right) \right].
   \label{omega-z-1PN}
\eea

For non-vanishing $v_x$ and $v_y$ (${\bf v}$ in ${\bf B_0}$-${\bf k}$ plane), Equations (\ref{x-comp-PN}) and (\ref{y-comp-PN}) give
\bea
   & & \left\{ {\omega^2 \over k^2}
       \left[ 1 + {1 \over c^2}
       \left( \Pi_0 + {p_0 \over \overline \varrho_0}
       \right) \right] \right\}^2
   \nonumber \\
   & & \qquad
       - {\omega^2 \over k^2}
       \left[ 1 + {1 \over c^2}
       \left( \Pi_0 + {p_0 \over \overline \varrho_0}
       \right) \right]
       \bigg\{
       c_s^2
       \left[ 1 - {1 \over c^2} \left( 4 U_0
       + 3 {4 \pi G \overline \varrho_0 \over k^2}
       + c_A^2 \right) \right]
       + c_A^2
       \left[ 1 - {1 \over c^2} \left( 6 U_0
       + {4 \pi G \overline \varrho_0 \over k^2}
       + c_A^2 \right) \right]
   \nonumber \\
   & & \qquad \qquad
       - {4 \pi G \overline \varrho_0 \over k^2}
       \left[ 1 + {1 \over c^2}
       \left( 2 \Pi_0 + 2 {p_0 \over \overline \varrho_0}
       - 2 U_0
       + 2 {4 \pi G \overline \varrho_0 \over k^2}
       - c_A^2 \right) \right]
       + {1 \over c^2} {({\bf k} \cdot {\bf n})^2 \over k^2}
       c_A^2 \left( c_s^2 + 4 {4 \pi G \overline \varrho_0 \over k^2}
       \right) \bigg\}
   \nonumber \\
   & & \qquad
       + {({\bf k} \cdot {\bf n})^2 \over k^2}
       c_A^2
       \left\{ c_s^2
       \left[ 1 - {1 \over c^2}
       \left( 10 U_0
       - {4 \pi G \overline \varrho_0 \over k^2}
       + c_A^2 \right) \right]
       - {4 \pi G \overline \varrho_0 \over k^2}
       \left[ 1 + {1 \over c^2}
       \left( 2 \Pi_0 + 2 {p_0 \over \overline \varrho_0}
       - 8 U_0 + 6 {4 \pi G \overline \varrho_0 \over k^2}
       - c_A^2 \right) \right] \right\}
   \nonumber \\
   & & \qquad
       = 0.
   \label{dispersion-PN}
\eea
{Using
\bea
   & & {\cal A} \equiv
       - c_s^2 \left( 4 U_0
       + 3 {4 \pi G \overline \varrho_0 \over k^2}
       + c_A^2 \right)
       - c_A^2 \left( 6 U_0
       + c_A^2 \right)
       - {4 \pi G \overline \varrho_0 \over k^2}
       \left( 2 \Pi_0 + 2 {p_0 \over \overline \varrho_0}
       - 2 U_0
       + 2 {4 \pi G \overline \varrho_0 \over k^2} \right)
   \nonumber \\
   & & \qquad
       + {({\bf k} \cdot {\bf n})^2 \over k^2}
       c_A^2 \left( c_s^2 + 4 {4 \pi G \overline \varrho_0 \over k^2}
       \right),
   \nonumber \\
   & & {\cal B} \equiv
       - c_s^2
       \left( 10 U_0
       - {4 \pi G \overline \varrho_0 \over k^2}
       + c_A^2 \right)
       - {4 \pi G \overline \varrho_0 \over k^2}
       \left( 2 \Pi_0 + 2 {p_0 \over \overline \varrho_0}
       - 8 U_0 + 6 {4 \pi G \overline \varrho_0 \over k^2}
       - c_A^2 \right),
\eea}
we can write it as
\bea
   & & W^2
       - W \left( c_s^2 + c_A^2
       - {4 \pi G \overline \varrho_0 \over k^2}
       + {1 \over c^2} {\cal A} \right)
       + {({\bf k} \cdot {\bf n})^2 \over k^2}
       c_A^2 \left( c_s^2
       - {4 \pi G \overline \varrho_0 \over k^2}
       + {1 \over c^2} {\cal B} \right)
       = 0,
\eea
with two solutions
\bea
   & & W \equiv {\omega^2 \over k^2}
       \left[ 1 + {1 \over c^2}
       \left( \Pi_0 + {p_0 \over \overline \varrho_0}
       \right) \right]
       = {1 \over 2} \Bigg\{
       c_s^2 + c_A^2
       - {4 \pi G \overline \varrho_0 \over k^2}
       + {1 \over c^2} {\cal A}
   \nonumber \\
   & & \qquad
       \pm \sqrt{ \left( c_s^2 + c_A^2
       - {4 \pi G \overline \varrho_0 \over k^2} \right)^2
       - 4 {({\bf k} \cdot {\bf n})^2 \over k^2}
       c_A^2
       \left( c_s^2
       - {4 \pi G \overline \varrho_0 \over k^2} \right)
       + {2 \over c^2} \left[ {\cal A}
       \left( c_s^2 + c_A^2
       - {4 \pi G \overline \varrho_0 \over k^2} \right)
       - 2 {\cal B} {({\bf k} \cdot {\bf n})^2 \over k^2}
       c_A^2 \right]} \Bigg\}.
   \label{omega-solution}
\eea
\end{widetext}

\begin{figure}[h]
\hspace{-.4cm} \vspace{-.0cm}
\includegraphics[width=80mm,angle=90]{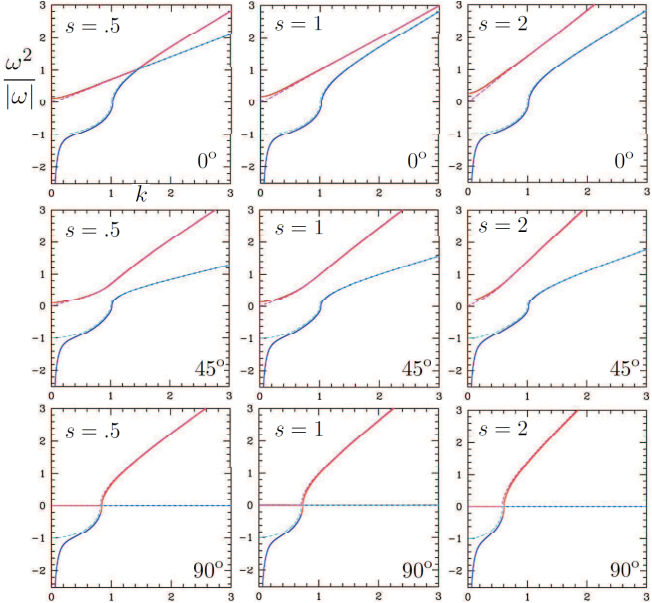}
\caption{{Behaviors of two 1PN solutions (solid lines) in
         Equation (\ref{omega-solution}) for $U_0 = 0$
         compared with the Newtonian ones (dashed lines) in
         Equation (\ref{omega-sol});
         the upper (plus) and lower (minus) signs
         correspond to red and blue lines for 1PN, and
         magenta and cyan lines for for Newtonian, respectively.
         We set $\Pi_0 = 0 = p_0$, and consider
         $R \equiv c_s^2 / c^2 = 0.01$,
         $s = 0.5$, $1$ and $2$, and $\psi = 0^{\rm o}$,
         $45^{\rm o}$ and $90^{\rm o}$.
         Units are the same as in Figure \ref{fig-BG}.}}
 \label{fig-pert}
\end{figure}

\begin{figure}[h]
\hspace{-.4cm} \vspace{-.0cm}
\includegraphics[width=80mm,angle=90]{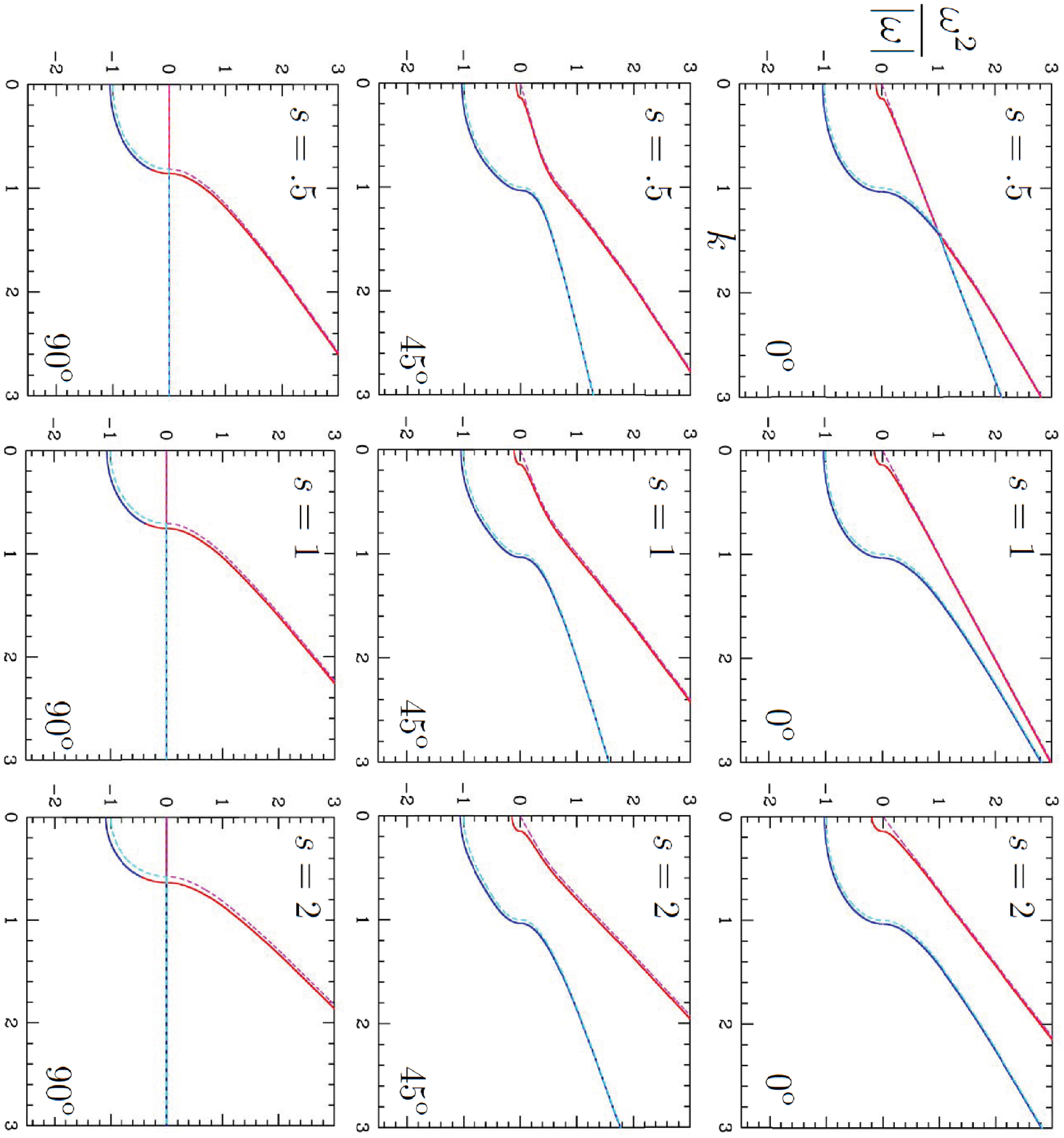}
\caption{{The same as Figure \ref{fig-pert}
         for $U_0 = 4 \pi G \overline \varrho_0/k^2$.}}
 \label{fig-pert-U}
\end{figure}

\begin{figure}[h]
\hspace{-.3cm} \vspace{-.0cm}
\includegraphics[width=65mm,angle=90]{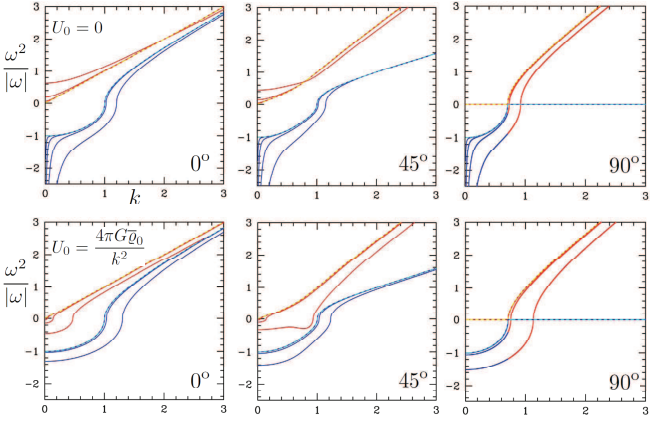}
\caption{{Dependence on $R \equiv c_s^2 / c^2$ for
         $R = 0.1$, $0.01$, $0.001$ and $0.0001$;
         the 1PN results (solid red and blue lines) approach
         the Newtonian ones (dashed yellow and cyan lines)
         as $R$ decreases.
         The upper and lower figures correxpond to
         $U_0 = 0$ and
         $U_0 = 4 \pi G \overline \varrho_0/k^2$, respectively.
         We consider $\psi = 0^{\rm o}$,
         $45^{\rm o}$ and $90^{\rm o}$ for
         $s \equiv c_A^2 / c_s^2 = 1$ and $\Pi_0 = 0 = p_0$.}}
 \label{fig-pert-R}
\end{figure}

{Behaviors of the 1PN solutions compared with
the Newtonian ones are presented in
Figures \ref{fig-pert} and \ref{fig-pert-U} for $U_0 = 0$
and $4 \pi G \overline \varrho_0/k^2$, respectively,
for a rather strong relativistic situation
with $R \equiv c_s^2 / c^2 = 0.01$.
Dependence on $R$ is presented in Figure \ref{fig-pert-R}.
As our analysis is valid to 1PN order the solutions in
Equation (\ref{omega-solution}) can be Taylor expanded to 1PN order.
In the Figures we plot these expanded solutions.}

{Notice that, in Figure \ref{fig-pert} for vanishing $U_0$,
the 1PN correction terms cause $\omega^2$ to diverge for small $k$ (large-scale) limit. For $k \rightarrow 0$, including $U_0$, we have
\bea
   & & {\omega^2 \over 4 \pi G \overline \varrho_0}
       \left[ 1 + {1 \over c^2}
       \left( \Pi_0 + {p_0 \over \overline \varrho_0}
       \right) \right]
   \nonumber \\
   & & \qquad
       = 0, \quad
       - 1 - 2 {4 \pi G \overline \varrho_0
       \over k^2 c^2} + 2 {U_0 \over c^2}.
   \label{k=0-limit}
\eea
Thus, for vanishing $U_0$, for small enough $k$,
the PN correction terms cause the system more unstable
and the exponent $i \omega$ diverges.
The diverging instability due to the 1PN correction
may imply breakdown of the 1PN approximation;
this was noticed by Nazari et al (2017) in the absence
of magnetic field. By demanding the 1PN correction terms
to be smaller than the 0PN order, we have
\bea
   & & k^2 > 2 {4 \pi G \overline \varrho_0 \over c^2}.
\eea
In terms of wavelength, $\lambda = 2 \pi/k$,
the 1PN approximation demands
\bea
   & & \lambda < c \sqrt{\pi \over 2 G \overline \varrho_0}
       \sim c t_g,
\eea
where $t_g \equiv 1/\sqrt{G \overline \varrho_0}$
is the gravitational timescale.
The diverging instability for $\lambda$ greater than
$c t_g$ (the light propagation distance during gravitational
timescale) is consistent with the eventual instability
of the background system [known and studied in cosmology
by Friedmann (1922, 1924) and Bonnor (1957)]
which we have ignored by {\it taking} the Minkowski background.
In a conservative stance, however, this simply implies
the limit of the 1PN approximation.}

{On the other hand, in Figure \ref{fig-pert-U}
for $U_0 = 4 \pi G \overline \varrho_0 / k^2$,
the 1PN correction terms in Equation (\ref{k=0-limit})
{\it disappear}. In the $k \rightarrow 0$ limit,
to the next leading order we have
\bea
   & & {\omega^2 \over 4 \pi G \overline \varrho_0}
       \left[ 1 + {1 \over c^2}
       \left( \Pi_0 + {p_0 \over \overline \varrho_0}
       \right) \right]
       =
       - {2 \over c^2} c_A^2 \cos^2{\psi},
   \nonumber \\
   & & \qquad
       - 1 - {1 \over c^2} \left( 7 c_s^2
       + 6 c_A^2 \sin^2{\psi} \right).
\eea}

For ${\bf k} \parallel {\bf B}_0$ ($\psi = 0^{\rm o}$), we have
\bea
   & & {\omega^2 \over k^2}
       \left[ 1 + {1 \over c^2}
       \left( \Pi_0 + {p_0 \over \overline \varrho_0}
       \right) \right]
   \nonumber \\
   & & \qquad
       = c_s^2 - {4 \pi G \overline \varrho_0 \over k^2}
       + {1 \over c^2} {\left( c_s^2
       - {4 \pi G \overline \varrho_0 \over k^2} \right) {\cal A}
       - c_A^2 {\cal B} \over c_s^2 - c_A^2
       - {4 \pi G \overline \varrho_0 \over k^2}},
   \nonumber \\
   & & \qquad
       c_A^2 + {1 \over c^2} {c_A^2 \left( - {\cal A}
       + {\cal B} \right) \over c_s^2 - c_A^2
       - {4 \pi G \overline \varrho_0 \over k^2}}.
\eea
One mode is unstable and the critical (Jeans) wavenumber becomes
\bea
   & & k_J^2 = {4 \pi G \overline \varrho_0 \over c_s^2}
       \left[ 1 + {1 \over c^2}
       \left( 2 \Pi_0 + 2 {p_0 \over \overline \varrho_0}
       + 5 c_s^2
       + 2 U_0 \right) \right].
   \label{k-J-PN}
\eea
This {\it coincides} exactly with the Jeans wavenumber in the absence of the magnetic field, see Equation (\ref{k-no-B}); thus, the criterion does not depend on the magnetic field as emphasized by Chandrasekhar \& Fermi (1953), now even to 1PN order, see below Equation (\ref{Jeans}). In the absence of gravity, we have
\bea
   & & {\omega^2 \over k^2}
       \left[ 1 + {1 \over c^2}
       \left( \Pi_0 + {p_0 \over \overline \varrho_0}
       \right) \right]
       = c_s^2, \quad
       c_A^2 \left( 1 - {c_A^2 \over c^2} \right).
   \label{omega-0-1PN}
\eea

\begin{widetext}
For ${\bf k} \perp {\bf B}_0$ ($\psi = 90^{\rm o}$), we have
\bea
   & & {\omega^2 \over k^2}
       = c_s^2
       \left[ 1 - {1 \over c^2}
       \left( \Pi_0 + {p_0 \over \overline \varrho_0}
       + c_A^2
       + 3 {4 \pi G \overline \varrho_0 \over k^2}
       + 4 U_0
       \right) \right]
       + c_A^2
       \left[ 1 - {1 \over c^2}
       \left( \Pi_0 + {p_0 \over \overline \varrho_0}
       + c_A^2
       + {4 \pi G \overline \varrho_0 \over k^2}
       + 6 U_0
       \right) \right]
   \nonumber \\
   & & \qquad
       - {4 \pi G \overline \varrho_0 \over k^2}
       \left[ 1 + {1 \over c^2}
       \left( \Pi_0 + {p_0 \over \overline \varrho_0}
       - c_A^2
       + 2 {4 \pi G \overline \varrho_0 \over k^2}
       - 2 U_0
       \right) \right], \quad
       0.
   \label{omega-90-1PN}
\eea
The critical wavenumber of the unstable mode becomes
\bea
   & & k_B^2 = {4 \pi G \overline \varrho_0 \over c_s^2 + c_A^2}
       \left[ 1 + {1 \over c^2}
       \left( 2 \Pi_0 + 2 {p_0 \over \overline \varrho_0}
       + 3 c_A^2 + 5 c_s^2
       + 2 U_0 {c_s^2 + 2 c_A^2 \over c_s^2 + c_A^2}
       \right) \right].
   \label{k-B-PN}
\eea

Now the choice for $U_0$ can be made.
We can either ignore the $U_0$ terms (Jeans' choice),
or use Equation (\ref{Poisson-BG}) to 0PN order,
thus $U_0 = 4 \pi G \overline \varrho_0/k^2$.
In the latter choice we have $U_0 = c_s^2$ in
Equation (\ref{k-J-PN}) and $U_0 = c_s^2 + c_A^2$ in
Equation (\ref{k-B-PN}).

Equations (\ref{k-J-PN}) and (\ref{k-B-PN}) show that
the 1PN corrections of the internal energy, pressure,
sound velocity and the Alfv\'en velocity cause to
{\it decrease} of critical (Jeans) wavelength,
$\lambda = 2 \pi/k$, thus {\it reduce} the Jeans mass,
$M \equiv (\pi/6) \overline \varrho \lambda^3$,
{Figures \ref{fig-pert}-\ref{fig-pert-R}
more generally show that for all $\psi$ values
the 1PN effects tend to increase
the critical wavenumber (thus lower the critical wavelength)
and make more negative $\omega^2$ (thus making system more unstable).
Therefore, we conclude that, to 1PN order,
all relativistic corrections tend to destabilize the system.}

\subsection{Limiting cases}

Here, we compare our results with previous studies.

(i) To 0PN order we recover Equation (\ref{0PN-dispersion}); Equation (169) in Chandrasekhar \& Fermi (1953).

(ii) Ignoring gravity, thus setting $U_0 \equiv 0 \equiv 4 \pi G \overline \varrho_0$, we recover the 1PN MHD waves in Section 7 of Hwang \& Noh (2020).

(iii) For vanishing magnetic field with ${\bf B} = 0$, thus setting ${\bf n} = 0 = c_A^2$, from Equation (\ref{v-2}) we have
\bea
   \bigg\{ {\omega^2 \over k^2}
       \left[ 1 + {1 \over c^2}
       \left( \Pi_0 + {p_0 \over \overline \varrho_0}
       + 3 {4 \pi G \overline \varrho_0 \over k^2}
       + 4 U_0
       \right) \right]
       - c_s^2
       + {4 \pi G \overline \varrho_0 \over k^2}
       \left[ 1 + {1 \over c^2}
       \left( 2 \Pi_0 + 2 {p_0 \over \overline \varrho_0}
       + 5 {4 \pi G \overline \varrho_0 \over k^2}
       + 2 U_0
       \right) \right]
       \bigg\} {\bf k} \cdot {\bf v}
       = 0.
\eea
\end{widetext}
By setting $\omega = 0$ we have $k = k_J$ with
\bea
   k_J^2 = {4 \pi G \overline \varrho_0 \over c_s^2}
       \left[ 1 + {1 \over c^2}
       \left( 2 \Pi_0 + 2 {p_0 \over \overline \varrho_0}
       + 5 c_s^2
       + 2 U_0 \right) \right].
   \label{k-no-B}
\eea
{This also follows from Equation (\ref{omega-solution}).}
By choosing $U_0 = 4 \pi G \overline \varrho_0 / k^2$,
we have $U_0 = c_s^2$ in Equation (\ref{k-no-B}).
{Therefore, the PN corrections of the internal energy,
pressure and sound velocity lower the critical (Jeans) wavelength.
Thus, as generally shown even in the presence of magnetic field,
all relativistic corrections tend to destabilize the system.}

Equation (\ref{k-no-B}), setting $U_0 = 0$, {\it differs}
from the result in Equation (56) of Nazari et al (2017)
in the coefficient of $c_s^2$. {In our approach of
separating the 0PN and 1PN orders clearly,} the difference
is caused by mixed decomposition of the 0PN and 1PN orders,
for example, in Equations (3) and (5) of Nazari et al (2017);
by moving the 1PN order terms in Equation (3) properly to
Equation (5) one can recover our result.

{In Nazari et al (2017)'s approach, however,
the 0PN and 1PN orders are {\it not} separated.
Compared to our $c_s^2$ based on $p = p(\overline \varrho)$,
thus $c_s^2 \equiv \delta p/ \delta \overline \varrho$,
their $c_s^2$ is based on $p = p(\varrho^*)$,
thus $c_s^2 \equiv \delta p/\delta \varrho^*$; $\varrho^*$
is the conserved density to 1PN order introduced by
Chandrasekhar (1965)
\bea
   & & \varrho^* \equiv  \overline \varrho \left[
       1 + {1 \over c^2} \left( {1 \over 2} v^2
       + 3 U \right) \right].
\eea
With this Equation (\ref{Mass-conserv-1PN}) can be written as
\bea
   & & {\partial \over \partial t} \varrho^*
       + \nabla \cdot \left( \varrho^* \overline {\bf v} \right)
       = 0.
   \label{PN-continuity}
\eea
Notice that $\varrho^*$ is already 1PN order,
and the same for $c_s^2$ defined based on $\varrho^*$.
Using this difference in the definition of $c_s^2$
we can also recover the $5$ factor in the coefficient of our $c_s^2$.
In this way the standard Jeans wavenumber in
Equations (26) and (35) of Nazari et al (2017)
already contains the 1PN order; similarly all results
in that paper are presented with the 0PN and 1PN orders
not clearly separated. Qualitatively, however,
as the coefficient of $c_s^2$ is changed from $2$ to $5$
without the sign change, the reducing effect on
the Jeans wavelength due to the PN sound velocity is not changed.
(We thank Professor Mahmood Roshan and Dr. Elham Nazari
for communications concerning their work.)}

\section{Stability of SR MHD}
                                    \label{sec:SR-stability}

We consider the same background medium with an aligned magnetic field. To the background order, we have a solution with
\bea
   & & \overline \varrho_0 = {\rm constant}, \quad
       \Pi_0 = {\rm constant}, \quad
       p_0 = {\rm constant},
   \nonumber \\
   & &
       {\bf B}_0 = B_0 {\bf n} = {\rm constant}, \quad
       {\bf v}_0 = 0, \quad
       \nabla \Phi_0 = 0,
\eea
but we have
\bea
   & & \Delta \Phi_0
       = 4 \pi G \left( \varrho_0 + {3 p_0 \over c^2}
       + {1 \over c^2} {1 \over 4 \pi} B_0^2 \right).
   \label{Poisson-BG-SR}
\eea
Thus, the inconsistency concerning $\Phi_0$ remains, but as in the Newtonian case $\Phi_0$ does not appear in the perturbation analysis, and we can ignore Equation (\ref{Poisson-BG-SR}), see a paragraph below Equation (\ref{0PN-BG}).

To the linear order, Equations (\ref{SR-MHD-eqs})-(\ref{eq4-SRG}) give
\bea
   & & {\partial \over \partial t}
       \left(
       \begin{array}{c}
           D   \\
           E   \\
           m^i \\
           B^i
       \end{array}
       \right)
       + \nabla_j
       \left(
       \begin{array}{c}
           D v^j  \\
           m^j c^2   \\
           m^{ij} \\
           v^j B^i - v^i B^j
       \end{array}
       \right)
       =
       \left(
       \begin{array}{c}
           0   \\
           0   \\
           - \overline \varrho \Phi^{,i} \\
           0
       \end{array}
       \right),
   \label{SR-MHD-eqs-linear} \\
   & & B^i_{\;\;,i} = 0,
   \\
   & & \Delta \Phi
       = 4 \pi G \left( \varrho + {3 p \over c^2}
       + {1 \over c^2} {1 \over 4 \pi} B^2 \right),
   \label{eq4-SRG-linear}
\eea
with
\bea
   & & D = \overline \varrho, \quad
       E/c^2 = \varrho
       + {1 \over c^2} {1 \over 8 \pi} B^2,
   \nonumber \\
   & & m^i = \left( \varrho + {p \over c^2} \right) v^i
       + {1 \over c^2} {1 \over 4 \pi} \left( B^2 v^i
       - B^i B^j v_j \right),
   \nonumber \\
   & & m^{ij} = p \delta^{ij}
       + {1 \over 4 \pi}
       \left( {1 \over 2} B^2 \delta^{ij}
       - B^i B^j \right).
   \label{D-def-linear}
\eea
Equations (\ref{eq2-SRG}) and (\ref{eq3-SRG}) are not needed.

Equations (\ref{SR-MHD-eqs-linear})-(\ref{eq4-SRG-linear}) can be arranged to give
\bea
   & & \dot \delta = - \nabla \cdot {\bf v},
   \label{Mass-conserv-SR} \\
   & & \delta \dot \Pi
       = - {p_0 \over \overline \varrho_0}
       \nabla \cdot {\bf v},
   \label{E-conserv-SR} \\
   & & \left[ 1 + {1 \over c^2} \left( \Pi_0
       + {p_0 \over \overline \varrho_0}
       + c_A^2 \right) \right]
       \dot {\bf v}
       - {1 \over c^2} c_A^2 {\bf n} {\bf n} \cdot \dot {\bf v}
   \nonumber \\
   & & \qquad
       + c_s^2 \nabla \delta
       + \nabla \delta \Phi
       = c_A^2 \left[
       {\bf n} \cdot \nabla {\bf b}
       - \nabla \left( {\bf n} \cdot {\bf b} \right)
       \right],
   \label{Mom-conserv-SR} \\
   & & \Delta \delta \Phi
       = 4 \pi G \overline \varrho_0
       \left[ \delta + {1 \over c^2}
       \left( \Pi_0 \delta + \delta \Pi
       + 3 c_s^2 \delta + 2 c_A^2 {\bf n} \cdot {\bf b} \right)
       \right],
   \nonumber \\
   \label{E-constr-SR} \\
   & & \dot {\bf b}
       = {\bf n} \cdot \nabla {\bf v}
       - {\bf n} \nabla \cdot {\bf v},
   \label{Maxwell-1-SR} \\
   & & \nabla \cdot {\bf b}
       = 0.
   \label{Maxwell-2-SR}
\eea
These can be compared with Equations (\ref{Mass-conserv-PN})-(\ref{Maxwell-2-PN}) in the PN case. As mentioned in a paragraph below Equation (\ref{eq3-SRG}), to be consistent we need $\delta \Phi$ only to the Newtonian order in Equation (\ref{E-constr-SR}); thus, $\Delta \delta \Phi = 4 \pi G \overline \varrho_0 \delta$; in the same spirit we ignore gravity terms combined with relativistic (PN) order, like $4 \pi G \overline \varrho_0 /(k^2 c^2)$. Thus, we do not need Equation (\ref{E-conserv-SR}). Compared with Newtonian Equations (\ref{Mass-conserv-0PN-linear})-(\ref{Maxwell-2-0PN-linear}) differences occur only in the coefficients of $\dot {\bf v}$ terms in Equation (\ref{Mom-conserv-SR}).

Expanding in plane waves we can derive
\begin{widetext}
\bea
   & & \left\{ \left[ 1 + {1 \over c^2} \left( \Pi_0
       + {p_0 \over \overline \varrho_0}
       + c_A^2 \right) \right]
       \omega^2
       - c_A^2 \left( {\bf n} \cdot {\bf k} \right)^2
       \right\}
       {\bf v}
       + \left[
       \left( {4 \pi G \overline \varrho_0 \over k^2}
       - c_s^2 - c_A^2 \right) {\bf k}
       + c_A^2 {\bf n} {\bf n} \cdot {\bf k}\right]
       {\bf k} \cdot {\bf v}
   \nonumber \\
   & & \qquad
       + c_A^2 \left(
       {\bf k} {\bf k} \cdot {\bf n}
       - {1 \over c^2} \omega^2 {\bf n} \right)
       {\bf n} \cdot {\bf v}
       = 0.
   \label{v-2-SR}
\eea
By taking a coordinate in Equation (\ref{coordinate}),
Equation (\ref{v-2-SR}) gives
\bea
   & & \bigg\{ \left[ 1 + {1 \over c^2}
       \left( \Pi_0 + {p_0 \over \overline \varrho_0}
       + c_A^2 \sin^2{\psi} \right) \right] \omega^2
       - k^2 \left( c_s^2 + c_A^2 \sin^2{\psi} \right)
       + 4 \pi G \overline \varrho_0 \bigg\} v_x
       + \left( k^2
       - {1 \over c^2} \omega^2 \right) c_A^2 \sin{\psi}
       \cos{\psi} v_y
       = 0,
   \label{x-comp-SR} \\
   & & \left( k^2
       - {1 \over c^2} \omega^2 \right) c_A^2 \sin{\psi}
       \cos{\psi} v_x
       + \bigg\{ \left[ 1 + {1 \over c^2}
       \left( \Pi_0 + {p_0 \over \overline \varrho_0}
       + c_A^2 \cos^2{\psi} \right) \right] \omega^2
       - c_A^2 k^2 \cos^2{\psi} \bigg\} v_y
       = 0,
   \label{y-comp-SR} \\
   & & \left\{
       \left[ 1 + {1 \over c^2}
       \left( \Pi_0 + {p_0 \over \overline \varrho_0}
       + c_A^2 \right) \right] \omega^2
       - c_A^2 k^2 \cos^2{\psi} \right\} v_z
       = 0.
   \label{z-comp-SR}
\eea
\end{widetext}

For $v_z \neq 0$ (${\bf v}$ perpendicular to ${\bf B}_0$-${\bf k}$ plane), Equation (\ref{z-comp-SR}) gives
\bea
   & & {\omega^2 \over k^2}
       = {c_A^2 \cos^2{\psi}
       \over 1 + {1 \over c^2}
       \left( \Pi_0 + {p_0 \over \overline \varrho_0}
       + c_A^2
       \right)}.
   \label{omega-z-SR}
\eea

For non-vanishing $v_x$ and $v_y$ (${\bf v}$ in ${\bf B_0}$-${\bf k}$ plane), from Equations (\ref{x-comp-SR}) and (\ref{y-comp-SR}) we have
\bea
   & & {\omega^4 \over k^4}
       \left[ 1 + {1 \over c^2}
       \left( \Pi_0 + {p_0 \over \overline \varrho_0}
       \right) \right]
       \left[ 1 + {1 \over c^2}
       \left( \Pi_0 + {p_0 \over \overline \varrho_0}
       + c_A^2 \right) \right]
   \nonumber \\
   & & \qquad
       - {\omega^2 \over k^2}
       \bigg\{
       \left( c_s^2 + c_A^2 \right)
       \left[ 1 + {1 \over c^2}
       \left( \Pi_0 + {p_0 \over \overline \varrho_0}
       \right) \right]
   \nonumber \\
   & & \qquad
       + {1 \over c^2} c_s^2 c_A^2 \cos^2{\psi}
       - {4 \pi G \overline \varrho_0 \over k^2} \bigg\}
   \nonumber \\
   & & \qquad
       + c_A^2 \cos^2{\psi} \left( c_s^2
       - {4 \pi G \overline \varrho_0 \over k^2} \right)
       = 0,
   \label{dispersion-SR}
\eea
with two solutions for $\omega^2/k^2$.

For ${\bf k} \parallel {\bf B}_0$ ($\psi = 0^{\rm o}$), we have
\bea
   & & {\omega^2 \over k^2}
       = { c_s^2 \over
       1 + {1 \over c^2}
       \left( \Pi_0 + {p_0 \over \overline \varrho_0} \right)}
       - {4 \pi G \overline \varrho_0 \over k^2}, \quad
   \nonumber \\
   & & \qquad
       {c_A^2 \over
       1 + {1 \over c^2}
       \left( \Pi_0 + {p_0 \over \overline \varrho_0}
       + c_A^2 \right)}.
   \label{omega-0-SR}
\eea
One mode is unstable and the critical (Jeans) wavenumber becomes
\bea
   & & k_J^2 = {4 \pi G \overline \varrho_0 \over c_s^2},
   \label{k-J-SR}
\eea
as the gravity part is valid only to the Newtonian order.

For ${\bf k} \perp {\bf B}_0$ ($\psi = 90^{\rm o}$), we have
\bea
   & & {\omega^2 \over k^2}
       = {c_s^2 + c_A^2 \over
       1 + {1 \over c^2}
       \left( \Pi_0 + {p_0 \over \overline \varrho_0}
       + c_A^2
       \right)}
       - {4 \pi G \overline \varrho_0 \over k^2}, \quad
       0.
   \label{omega-90-SR}
\eea
The critical wavenumber of the unstable mode similarly becomes
\bea
   & & k_B^2 = {4 \pi G \overline \varrho_0 \over c_s^2 + c_A^2}.
   \label{k-B-SR}
\eea

As mentioned, as we consider the gravity the stability
analysis is consistent only to the Newtonian order,
thus Equations (\ref{k-J-SR}) and (\ref{k-B-SR})
are the same as in the Newtonian MHD.
Whereas, considering gravity only to the Newtonian order,
the wave solutions in Equations (\ref{omega-z-SR}),
(\ref{omega-0-SR}) and (\ref{omega-90-SR}),
and the two general solutions of Equation (\ref{dispersion-SR})
for general $\psi$ are valid for the fully SR MHD.
In the absence of gravity, the wave speeds in
Equations (\ref{omega-z-SR}), (\ref{omega-0-SR})
and (\ref{omega-90-SR}) include the 1PN results in
Equations (\ref{omega-z-1PN}), (\ref{omega-0-1PN})
and (\ref{omega-90-1PN}) as the 1PN limit.

\section{Summary}
                                    \label{sec:Summary}

Sections \ref{sec:PN-MHD} and \ref{sec:SR-MHD} are summaries
of two formulations of relativistic MHD: MHD valid to
1PN approximation and the SR MHD with weak gravity.
The equations of PN MHD are presented without taking
the slicing condition, whereas the equations of SR MHD
with weak gravity are valid in the maximal slicing which
corresponds to the harmonic gauge in the PN formulation.
Section \ref{sec:0PN-stability} is the MHD instability of
the the homogeneous medium with an aligned magnetic field
in Newtonian (0PN) limit, largely overlapping with previous
studies in Chandrasekhar \& Fermi (1954) and
Chandrasekhar (1954, 1961). Here, we clarify some unclear
remarks made in the previous works concerning effects
of magnetic field on the gravitational instability,
see below our Equations (\ref{Jeans}) and (\ref{Jeans-B}).
We also address the inconsistency issue related to the
Poisson's equation to the homogeneous background medium,
see paragraphs below Equation (\ref{0PN-BG}) and below
Equation (\ref{Gauge-PN}).

Section \ref{sec:1PN-stability} presents our main analysis
of the 1PN gravito-magnetic instability of the same medium.
We show that the post-Newtonian corrections of the internal
energy ($\Pi_0$), pressure ($p_0$), sound velocity ($c_s$)
and the Alfv\'en velocity ($c_A$) consistently lower the
Jeans wavelength and the Jeans mass, see Equations
(\ref{k-J-PN}) and (\ref{k-B-PN}),
{and tend to destabilize the system,
see Figures \ref{fig-pert}-\ref{fig-pert-R}.}
We note that we have not
fixed the temporal gauge condition for the analysis,
thus the results are valid independently of the gauge
condition and are naturally gauge invariant,
see Section 6.3 in Hwang \& Noh (2020).

Section \ref{sec:SR-stability} presents another main
analysis of the waves and instability of the SR MHD
with weak gravity. The MHD waves of the SR MHD are
presented in Equations (\ref{omega-z-SR})-(\ref{omega-0-SR})
and (\ref{omega-90-SR}). When we consider the gravity however,
the analysis is consistent to Newtonian order only,
and the critical wavenumbers in Equations (\ref{k-J-SR})
and (\ref{k-B-SR}) are the same as in the Newtonian MHD.

\section{Discussion}
                                    \label{sec:Discussion}

{Stability analysis of an infinite homogeneous medium
is a textbook exercise in the Newtonian case;
gravitational instability in the presence of aligned
magnetic field is rarely presented though (Chandrasekhar 1961).
The exponential instability for an imaginary $\omega$
indicates no lack-of-time problem in the gravitational collapse
in a static medium: i.e., the over-dense region larger than
Jeans scale may collapse immediately.
This may have a negative side though that
we can hardly find astrophysical situations where
the conditions (static homogeneous medium, let aside the
infinity condition) are met in the interstellar space.
Interstellar matter and molecular clouds, in fact,
are known to be in the state of (compressible) turbulence
(Elmegreen \& Scalo 2004).
}

{Although the infinite static homogeneous background medium
has the internal inconsistency (which remains even in
the relativistic situation as we have shown in this work)
the Jeans stability criterion survives
in a consistent relativistic analysis made in infinite homogeneous,
but {\it dynamic}, background medium (Lifshitz 1946);
in the static limit we recover the Jeans result!
In a power-law expanding medium (like in the matter dominated era)
the growth rate is suppressed from exponential
to a decelerating power-law; this was originally recognized
as too slow to serve for galaxy or star formation by Lifshitz (1946).
Nowadays, our observable universe provides a prime example
with highly successful gravitational instability
as the main driving engine of the large-scale structure
with a long duration (due to slow growth rate)
from its generation in the early universe
till even today.
}

{Lifshitz's stability analysis, however,
is made in the absence of magnetic field.
In the presence of aligned magnetic field, the proper relativistic
analysis demands other than the Friedmann cosmology
with the anisotropic cosmological principle (Thorne 1967).
Current success in cosmology does not encourage such a study,
and which part of our study could survive the consistent analysis
(in the static limit) is currently unknown.
}

{As the static homogeneous medium is not realistic
(i.e, cannot or rarely be found) in known astrophysical environments,
an additional aligned magnetic field is also apparently not realistic.
Our motivation for choosing the situation is to extend
the textbook analysis of the gravito-magnetic instability
known in Newtonian MHD to a couple of relativistic situations.
We have shown that the inconsistency noticed by Jeans is
due to the static nature of the background medium.
The relativistic analysis can cure the inconsistency by demanding
{\it dynamic} background medium
(Lifshitz 1946; Hwang, Noh \& Puetzfeld 2008).
}

{More realistic astrophysical situations where
relativistic MHD is needed can be found
in accretion discs, magnetospheres, the plasma winds
and astrophysical jets near compact astrophysical objects
(like neutron stars and black holes), and active galactic nuclei.
Such situations may require the spherical, cylindrical or
disc geometries (Thorne, Price \& Macdonald 1986; Beskin 2010).
Stability analysis in these weakly relativistic situations
using our two new relativistic MHD formulations
summarized in Sections \ref{sec:PN-MHD} and \ref{sec:SR-MHD}
will be interesting.
Being fully nonlinear, these two relativistic MHD formulation
may be convenient to numerically handle the weakly relativistic
(1PN approximation) or weak gravity situations.
}

%
%
\section*{Acknowledgments}
{We wishes to thank the referee for constructive comments.}
J.H.\ was supported by Basic Science Research Program
through the National Research Foundation (NRF) of Korea
funded by the Ministry of Science, ICT and future
Planning (No.\ 2018R1A6A1A06024970 and NRF-2019R1A2C1003031).
H.N.\ was supported by National Research Foundation of Korea
funded by the Korean Government (No.\ 2018R1A2B6002466).

%
%

\end{document}